\documentclass[12pt,a4]{article}

\usepackage{color,tikz}

\usepackage{amsmath,amssymb,amsthm,amsfonts}
\usepackage{mathrsfs,graphicx,texdraw,xcolor}

\newtheorem{remark}{Remark}

\def\neprod{\setbox0=\hbox{$\nearrow$}
  \box0\kern-1.6em\prod} 
\def\swprod{\setbox0=\hbox{$\swarrow$}%
  \,\,\raise.03em\box0\kern-1.18em\prod} 
\def\seprod{\setbox0=\hbox{$\searrow$}%
  \,\,\raise.00em\box0\kern-2.0em\prod} 

\def\openone{\leavevmode\hbox{\small1\kern-3.3pt\normalsize1}}

\def\a{{\boldsymbol a}}
\def\b{{\boldsymbol b}}
\def\m{{\boldsymbol m}}

\def\n{{\boldsymbol n}}

\def\u{{\boldsymbol u}}
\def\v{{\boldsymbol v}}

\def\s{{\boldsymbol s}}
\def\p{{\boldsymbol p}}
\def\q{{\boldsymbol q}}

\begin{document}

\begin{center}
{\LARGE \bf Multicomponent  Fokas-Lenells equations on Hermitian symmetric spaces}

\bigskip

{\bf Vladimir S. Gerdjikov$^{a,b,c,}$ \footnote{E-mail: {\tt gerjikov@inrne.bas.bg}} and
Rossen I. Ivanov$^{d,}$ \footnote{E-mail: {\tt rossen.ivanov@tudublin.ie}}
}

\medskip

\noindent
{\it $^a$  Department of Applied Mathematics,  National Research Nuclear University MEPHI, \\
31 Kashirskoe Shosse, 115409 Moscow, Russian Federation}\\[5pt]
{\it $^b$ Institute of Mathematics and Informatics, Bulgarian Academy of Sciences,\\
8 Georgi Bonchev Str.,  1113 Sofia, Bulgaria}\\[5pt]
{\it $^c$ Institute for Advanced Physical Studies, New Bulgarian University,\\ 21 Montevideo Str. Sofia 1618, Bulgaria}\\[5pt]
{\it $^d$ School of Mathematical Sciences, Technological University Dublin, City Campus, \\  Kevin Street, Dublin, D08 NF82, Ireland}
\\[25pt]

\end{center}

{\small{\bf  Abstract.} Multi-component integrable generalizations of the Fokas-Lenells equation, associated with each irreducible Hermitian symmetric space are formulated. Description of the underlying structures associated to the integrability, such as the Lax representation and the bi-Hamiltonian formulation of the equations is provided. Two reductions are considered as well, one of which leads to a nonlocal integrable model. Examples with Hermitian symmetric spaces of all classical series of types A.III, BD.I, C.I and D.III are presented in details, as well as possibilities for further reductions in a general form.
\\[5pt]
}

\section{Introduction}
The Fokas-Lenells  (FL) equation introduced in \cite{Fo} and studied further in \cite{LeFo, LeFo2, Le} has been at the center of a considerable amount of research in the recent years. The FL equation bears a resemblance to the well-known integrable Nonlinear Schr\"odinger Equation (NLS) \cite{FaTa, GVY} and the Derivative NLS (DNLS or DNLS I)
\cite{KN, GIK, For} as well as DNLS~II \cite{ChLi} and DNLS~III {\cite{GI0, GI} equations. It shares some features with other integrable equations in non-evolutional form such as the Camassa-Holm (CH) equation\footnote{The CH equation was found after its derivation as a shallow water equation in \cite{CaHo} to fit into a class of integrable equations derived previously by using hereditary symmetries in \cite{FF1, FF2}.} such as negative powers of the spectral parameter in the $M$-operator in the Lax representation, i.e. related to the ``negative'' flows of the corresponding hierarchy of integrable equations. The interest to the equations from the ``negative''  flows is to a big extent related to the variety and complexity of their solutions, \cite{LeFo, Le, Ma1, Ma2, CaHo, HoIv, HeXuPor}.  Multi-component generalizations of the FL equation have appeared recently in numerous studies like \cite{89, VG-tmf92, Guo, HeXuPor, LiFeZhu, YZ, Ye, ZhHe, ZYC, Ma3} and this naturally leads to the need of their classification from the viewpoint of the simple Lie algebras, the associated symmetric spaces and their reductions. The other multi-component integrable equations in a non-evolutionary form include for example the massive Thirring-like model, whose integrability was shown by Kuznetsov and Mikhailov \cite{KuMi}; its multicomponent extensions were proposed in \cite{Tsu}.

Here we study generalizations of the FL equation with possible reductions by considering integrable systems associated to some simple Lie algebra $\frak{g}$ over the complex numbers and their Hermitian symmetric spaces.
 The structure of a symmetric space is determined by an involutive automorphism of the Lie algebra $\frak{g}$, known as Cartan involution. There is a decomposition of the Lie algebra \begin{align*}
  \mathfrak g= \mathfrak{g}^{(0)}\oplus \mathfrak{g}^{(1)}\, ,
\end{align*}
  where $\mathfrak{g}^{(0)}$ is a subalgebra, invariant under the Cartan involution, and  $\mathfrak{g}^{(1)}$ is a complementary subspace on which the Cartan involution has an eigenvalue $-1$. The orthogonality between $\mathfrak{g}^{(0)}$ and $\mathfrak{g^{(1)}}$ is with respect to the Killing form of $\mathfrak{g}.$ 

The classification of the symmetric spaces of the simple Lie groups is provided in the classic monograph \cite{h}. The Hermitian symmetric spaces form a special subclass and their classification could also be found in \cite{h}.
Due to the Lie-algebraic nature of this splitting, the Hamiltonian variables of these equations separate into sets taking values in either $\mathfrak{g}^{(0)}$ or $\mathfrak{g}^{(1)}$.
Moreover, by restricting the Hamiltonian to depend only on the variables in the space $\mathfrak{g}^{(1)}$, the arising integrable nonlinear equations can be written in terms of variables taking values in a symmetric space.

Integrable systems on symmetric spaces of finite dimensional Lie algebras have been studied considerably in the literature, see \cite{KuFo, AtFo,For,Basic, GeGr, GGK, ArHoIv}.

The paper is organized as follows. In section \ref{s2} we provide some preliminary facts from the theory of the simple Lie algebras and Hermitian symmetric spaces, which are necessary to introduce the relevant Lax representations. In Section \ref{s4} we derive the generalized equations by using Lax operators with values in an irreducible Hermitian symmetric space. Two straightforward reductions are given as well, one of them leads to a {\it nonlocal} nonlinear integrable equation, which in addition depends on the {\it reflected} independent variables $(-x,-t)$. The reductions are closely related to the discrete symmetries of the equations, including symmetries involving space and time reflections \cite{Mi, V, GeGrIv, Gu1, Gu2}.   The Hamiltonian structures are discussed in section \ref{s5}. Four examples for specific choices of a symmetric space are given in Section \ref{s6}. Section \ref{s7} contains some additional types of reductions reductions of these Fokas-Lennells equations. The last Section \ref{s8} contains conclusions and discussions.

\section{Preliminaries}\label{s2}
We assume that the readers are familiar with the theory of simple Lie algebras and with the basics of differential geometry.

\subsection{Simple Lie algebras and Cartan-Weyl generators }
Let $\mathfrak{g}$ be simple Lie algebra and let $\Delta $ be its root system. Here we fix up the notations and the normalization conditions for the
Cartan-Weyl generators of $\mathfrak{g}$. The commutation relations are given by \cite{LA, Goto}:
\begin{equation}\label{eq:CW0}\begin{split}
[H_{e_{k}},E_{\alpha }] &=(\alpha ,e_{k})E_{\alpha },\\ [E_{\alpha}, E_{-\alpha }]&=H_{\alpha } =\sum_{k=1}^{r}(\alpha ,e_{k})H_{e_k } \\
[E_{\alpha },E_{\beta }] &=\left\{\begin{array}{ll} N_{\alpha ,\beta }E_{\alpha +\beta }\quad & \;\alpha +\beta \in \Delta \\ 0 & \;\alpha +\beta \notin\Delta .\end{array}\right.
\end{split} \end{equation}
where $\Delta $ is the root system of $\mathfrak{g}$, $H_{e_{k}}$,
$k=1,...,r$ ($r=\mathrm{rank}\,\mathfrak{g}$) is the basis of the Cartan subalgebra $\mathfrak{h}$ of $\mathfrak{g}$ and $E_{\alpha }$ are the remaining elements of the Cartan-Weyl basis associated to each root $\alpha \in \Delta.$  Here and below $e_{k}$, $\alpha $ are vectors in an $r$-dimensional Euclidean space\footnote{For the algebras of type ${\bf A_r}$ such as $sl(r+1,\mathbb{C}),$  $su(r+1, \mathbb{C})$ which are of rank $r$, the Euclidean root space is $(r+1)$-dimensional, while the root system spans the $r$-dimensional hyperplane orthogonal to the vector $e_1+e_2+\cdots + e_{r+1}$.}, associated to the Cartan elements $H_{e_{k}}$, $H_{\alpha }$ correspondingly.
The Euclidean inner product is denoted by $(\cdot, \cdot)$.
The quantities $N_{\alpha ,\beta }$ have various properties such as obviously $N_{\beta ,\alpha }=-N_{\alpha ,\beta }$, see \cite{LA, Goto}. The Killing form provides a metric on $\mathfrak{g}$ and is defined as
$$\langle X, Y \rangle = \text{tr}(\text{ad}_X \text{ad}_Y)$$ where $\text{ad}_X,$ $\text{ad}_Y $  are the elements $X, Y\in \mathfrak{g}$ taken in the adjoint representation. Since $\mathfrak{g}$ is simple, the Killing form is proportional to the trace taken in any irreducible representation (say the fundamental representation), $$\langle X, Y \rangle = K\, \text{tr}(XY)$$ for some constant $K$. Indeed, there is a homomorphism between the two representations and since the adjoint representation is irreducible when $\mathfrak{g}$ is simple then the homomorphism is isomorphism by Schur's lemma.   The normalization of the basis is determined by the Killing form such that
\begin{equation}
E_{-\alpha }=E_{\alpha }^{T},\quad \langle E_{-\alpha },E_{\beta }\rangle =  \delta^{\alpha}_{ \beta},  \nonumber
\end{equation} where $\delta$ is the Kronecker's symbol. On the other hand \begin{equation}\label{eq:KF}
\langle E_{\alpha}, E_{\beta}\rangle =  \delta_{\alpha+\beta, 0}
\end{equation} and it is useful to introduce a metric tensor on   $\mathfrak{g}/ \mathfrak{h}$  \begin{equation} \label{g} g_{\alpha, \beta}=\langle E_{\alpha}, E_{\beta} \rangle =\delta_{\alpha, -\beta}, \qquad \alpha, \beta \in \Delta. \end{equation}

We recall also that if $\alpha$ is a root, then $-\alpha$ is also a root of $\mathfrak{g}$. Thus the root system can be split into sets of positive and negative roots $\Delta = \Delta^+ \cup \Delta^-$. The canonical way of introducing the root systems for all simple Lie algebras is well known \cite{h}.

\subsection{Hermitian Symmetric spaces}\label{s3}

The symmetric spaces are associated to a Cartan involution acting on the corresponding Lie group elements. This involution has an  induced action $\varphi$ on  $\mathfrak{g}$ such that
\begin{align}
 &\mathfrak{g}^{(0)} \equiv \{ X \in \mathfrak{g} \, | \quad \varphi(X)=X  \}\\
&\mathfrak{g}^{(1)} \equiv \{ X \in \mathfrak{g} \, | \quad \varphi(X)=-X  \}
\end{align}
This way the Cartan involution (the automorphism $\varphi$) introduces a $\mathbb{Z}_2$-grading on $\mathfrak{g}$, i.e.
\begin{equation}\label{eq:g0}\begin{split}
 \mathfrak{g} = \mathfrak{g}^{(0)} \oplus \mathfrak{g}^{(1)}
\end{split}\end{equation} such that
\begin{align}
        [\mathfrak{g}^{(0)}, \mathfrak{g}^{(0)}] \subset \mathfrak{g}^{(0)}\, , \qquad [\mathfrak{g}^{(0)},\mathfrak{g}^{(1)} ] \subset  \mathfrak{g}^{(1)}\, , \qquad [\mathfrak{g}^{(1)},\mathfrak{g}^{(1)}] \subset \mathfrak{g}^{(0)}\, .
    \label{sym-relation}
\end{align}

In addition, $\mathfrak{g}^{(0)} $ is a subalgebra of $\mathfrak{g}.$ Denoting by $K$ and $G $ the Lie groups, associated to $\mathfrak{g}^{(0)} $ and $\mathfrak{g}$ correspondingly, the linear subspace $\mathfrak{g}^{(1)} $ is  identified with the tangent space of $G/K,$ which is used as a notation for the corresponding symmetric space.

The Hermitian symmetric spaces are a special class of symmetric spaces for which the Cartan involution $\varphi$ is related to a special element $J\in \mathfrak{h} $ such that

\begin{description}

  \item[(i)] The Lie sub-algebra $\mathfrak{g}^{(0)}$ is \begin{equation}\label{eq:g01}\begin{split}
 \mathfrak{g}^{(0)} \equiv \{ X \in \mathfrak{g} \, | \quad [J, X] =0 \} ,
\end{split}\end{equation}
i.e. $X = \sum_{k=1}^{r} x_k H_{e_k} + \sum_{\beta\in \Delta_0} x_\beta E_\beta$;
$\mathfrak{g}^{(1)}$ is the vector space complement of $\mathfrak{g}^{(0)}$ in  $\mathfrak{g}.$

\item[(ii)] The root system $\Delta$ could be decomposed into two sets, such that $\alpha(J)\equiv \vec{\alpha}\cdot \vec{J}$ takes integer values $0$ or $\pm a, $ $(a>0)$ for all $\alpha\in \Delta$ on each set:
\begin{equation}\label{eq:g02}\begin{split}
 \Delta = \Delta_0 \cup \Delta_1, \qquad \Delta_0 \equiv \{ \alpha \in \Delta \quad \mbox{such that} \quad \alpha(J) =0 \},
\end{split}\end{equation}
\begin{equation}\label{eq:roots2}\begin{split}
 \Delta _1 = \Delta_1 ^{+}\cup \Delta_1 ^{-}, \qquad \Delta_1^{\pm} \equiv \{ \alpha \in \Delta_1 \quad \mbox{such that} \quad \alpha(J) =\pm a , \quad a>0 \, \}. \end{split}\end{equation}

Note that
\begin{equation}\label{eq:com1}
[J, E_{\alpha}]=\alpha(J) E_{\alpha}=\pm a E_{\alpha}, \qquad   \alpha \in \Delta _1 ^{\pm}
\end{equation}
and $a>0$ is a constant for the selected Hermitian symmetric space.

\item[(iii)] $[E_{\alpha}, E_{\beta}] =0 $ if both $\alpha, \beta \in \Delta_1^{+}$ or $\alpha, \beta \in \Delta_1^{-},$  this follows from (ii).

The classification of the irreducible symmetric spaces and the subclass of irreducible Hermitian symmetric spaces is provided for example in \cite{h}.
\end{description}

We need the following quantities $$R_{\alpha,\beta, \gamma, \delta}=\langle [ E_{\alpha}, E_{\beta}], [ E_{\gamma}, E_{\delta}]\rangle .$$ By its definition it has all the symmetries of the Riemann tensor. With the definition of the metric tensor we have also
$$R_{\alpha,\beta, \gamma, \delta}=g_{\alpha,\lambda}R^{\lambda}_{\phantom{*}\beta, \gamma, \delta}=\delta_{\alpha+\lambda,0}R^{\lambda}_{\phantom{*}\beta, \gamma, \delta}=R^{-\alpha}_{\phantom{***}\beta, \gamma, \delta}.$$

Let us suppose that $E_{\alpha}$ are taken in a matrix representation where all their matrix entries are real. Then we have the following properties:

1. $R_{\alpha,\beta, \gamma, \delta}=R_{-\alpha,-\beta, -\gamma, -\delta}$. It follows from the properties of the trace:
$$R_{\alpha,\beta, \gamma, \delta}=K \, \text{tr} ([ E_{\alpha}, E_{\beta}] [ E_{\gamma}, E_{\delta}])= K \, \text{tr} ([ E_{\alpha}, E_{\beta}]^T [ E_{\gamma}, E_{\delta}]^T)=K \, \text{tr} ([ E_{-\alpha}, E_{-\beta}] [ E_{-\gamma}, E_{-\delta}]).$$


2. Suppose that $\alpha, \beta,\gamma, \delta \in \Delta_1^+.$ Then
\begin{equation} \label{pr2}
R_{-\alpha, \gamma,-\beta, \delta}=R_{-\alpha, \delta,-\beta,\gamma}.
\end{equation}

Proof: Using the properties of the trace after expanding the commutators, we have
\begin{align}
&R_{-\alpha, \gamma,-\beta, \delta}-R_{-\alpha, \delta,-\beta,\gamma}=
K \, \text{tr} ([ E_{-\alpha}, E_{\gamma}] [ E_{-\beta}, E_{\delta}])-K \, \text{tr} ([ E_{-\alpha}, E_{\delta}] [ E_{-\beta}, E_{\gamma}]) \nonumber \\ \nonumber
&=K \, \text{tr} ([ E_{-\alpha}, E_{-\beta}] [ E_{\gamma}, E_{\delta}])=0
\end{align}
since both commutators in the last expression are zero, due to property (iii) above.

\subsection{Generic form of Lax representations}

Here we outline the generic form of Lax representations which are polynomial in the spectral parameter $\lambda$ and are compatible with the structure of the symmetric space $G/K$.

The simplest nontrivial classes of such Lax operators were introduced by Fordy  and Kulish \cite{KuFo}. The first one is linear in $\lambda$
\begin{equation}\label{eq:La1}\begin{split}
L_1\psi \equiv i \frac{\partial \psi}{ \partial x } + (\mathcal{Q}(x,t) - \lambda J)\psi(x,t,\lambda) =0
\end{split}\end{equation}
and generates the class of multicomponent NLS equations. This case has been very well studied, so we pay more attention to the second one, which is quadratic in $\lambda,$
\begin{equation}\label{eq:La2}\begin{split}
L_2\psi \equiv i \frac{\partial \psi}{ \partial x } + (\lambda \mathcal{ Q}(x,t) - \lambda^2 J)\psi(x,t,\lambda) =0
\end{split}\end{equation}
generates the class of multicomponent derivative NLS equations. The Lax operators generating the class of GI equations \cite{GI0, GI} and the class of Chen-Lie-Liu equations \cite{ChLi} are related to (\ref{eq:La2}) by simple gauge transformations.

In both cases the gauge is fixed by choosing the leading term in $\lambda$ to be constant diagonal matrix $J$ which determines the Cartan involution.  Next, the potential $\mathcal{Q}(x,t) = [J, \tilde{Q}(x,t)]$, where $\tilde{Q}(x,t)$ is a generic element of $\mathfrak{g}$, then
\begin{equation}\label{eq:Qdef}
\mathcal{Q}(x,t)=\sum_{\alpha \in \Delta^+_1} (q^{\alpha} E_{\alpha}+p^{\alpha} E_{-\alpha}).
\end{equation}
Thus $\mathcal{Q}(x,t) \in \mathfrak{g}^{(1)}$ in fact determines the local coordinates in the tangent space of $G/K$. The coefficients $q_\alpha$ and $p_\alpha$ can be evaluated by using the Killing form:
\begin{equation}\label{eq:}\begin{split}
q_\alpha (x,t) = \langle\mathcal{ Q}(x,t) E_{-\alpha}\rangle , \qquad p_\alpha (x,t) = \langle\mathcal{ Q}(x,t) E_{\alpha}\rangle .
\end{split}\end{equation}
In what follows we assume that $q_\alpha$ and $p_\alpha$ are smooth functions of $x$ and $t$ tending to 0 for $|x|\to \infty$.

\section{Fokas-Lenells equation on Hermitian symmetric spaces} \label{s4}

The Fokas-Lenells equations are associated to the so-called {\it negative flows} and the following Lax pair
\begin{equation}\label{eq:Lax}\begin{split}
& i \Psi _x + (\lambda Q_x - \lambda^2 J) \Psi =0,\\
& i \Psi_t +\left(\lambda Q_x  + V_0 + \lambda^{-1} V_{-1} -(\lambda^2-\frac{2}{a}+ \frac{1}{a^2\lambda^{2}}) J\right) \Psi,
\end{split}\end{equation} where
\begin{equation}\label{eq:Q}
Q(x,t)=\sum_{\alpha \in \Delta^+_1} (q^{\alpha} E_{\alpha}+p^{\alpha} E_{-\alpha}).
\end{equation}
From the compatibility condition $(i \Psi _x)_t-(i \Psi _t)_x=0$ and the requirement that it should be satisfied identically for any value of the spectral parameter $\lambda$ we obtain the following equations as coefficients in the expansion of the compatibility condition in powers of $\lambda:$
\begin{equation}\label{eq:gen}\begin{aligned}
& \lambda: &\quad \qquad & -i \frac{\partial^2 Q}{ \partial x \partial t } + i \frac{\partial^2 Q}{ \partial x^2} + \left[ \frac{\partial Q}{ \partial x }, V_0 \right] -[J,V_{-1}]  +\frac{2}{a} \left[ \frac{\partial Q}{ \partial x },J \right]=0, \\
& \lambda^0: &\quad \qquad & i \frac{\partial V_0}{ \partial x } + \left[ \frac{\partial Q }{ \partial x } ,V_{-1} \right]=0, \\
& \lambda^{-1}: &\quad \qquad & i \frac{\partial V_{-1}}{ \partial x } -\frac{1}{a^2} \left[ \frac{\partial Q}{ \partial x } ,J \right]=0.
\end{aligned}\end{equation}

The last two equations could be solved directly, yielding
\begin{equation}\label{eq:V0Vm1}\begin{aligned}
&V_{-1}=\frac{i}{a} \sum_{\alpha \in \Delta^+_1} (q^{\alpha} E_{\alpha}-p^{\alpha} E_{-\alpha})   , \\
& V_0  =\frac{1}{a} \sum_{\alpha, \beta \in \Delta^+_1} q^{\alpha}p^{\beta}[ E_{\alpha}, E_{-\beta}]  .
\end{aligned}\end{equation}
Following the analysis in \cite{KuFo}  the remaining equation for $Q$ gives
\begin{equation}\label{eq:Qa}\begin{aligned}
& \sum_{\alpha \in \Delta^+_1} \left(iq^{\alpha}_{xt}- iq^{\alpha}_{xx}  +i q^{\alpha} +2 q^{\alpha}_x\right)E_{\alpha}-\frac{1}{a} \sum_{\beta, \gamma, \delta \in \Delta^+_1} q^{\gamma}_x q^{\delta} p^{\beta} [E_{\gamma}, [ E_{\delta}, E_{-\beta}]]=0   , \\
& \sum_{\alpha \in \Delta^+_1} \left(ip^{\alpha}_{xt}- ip^{\alpha}_{xx}  +i p^{\alpha} - 2p^{\alpha}_x \right)E_{-\alpha}-\frac{1}{a} \sum_{\beta, \gamma, \delta \in \Delta^+_1} p^{\gamma}_x q^{\delta} p^{\beta} [E_{-\gamma}, [ E_{\delta}, E_{-\beta}]]=0    .
\end{aligned}\end{equation}
Using pairing with the Killing form, $\langle E_{-\alpha}, \phantom{*}\cdot \phantom{*}  \rangle $ for the first equation and $\langle E_{\alpha}, \phantom{*}\cdot \phantom{*}  \rangle $ for the second one we have:
\begin{equation}\label{eq:Q2}\begin{aligned}
& iq^{\alpha}_{xt}- iq^{\alpha}_{xx}  +i q^{\alpha}  +2 q^{\alpha}_x -\frac{1}{a} \sum_{\beta, \gamma, \delta \in \Delta^+_1}
q^{\gamma}_x q^{\delta} p^{\beta} \langle E_{-\alpha}, [E_{\gamma}, [ E_{\delta}, E_{-\beta}]]\rangle =0   , \\
&ip^{\alpha}_{xt}- ip^{\alpha}_{xx} +ip^{\alpha} -2 p^{\alpha}_x -\frac{1}{a} \sum_{\beta, \gamma, \delta \in \Delta^+_1}
p^{\gamma}_x q^{\delta} p^{\beta} \langle E_{\alpha}, [E_{-\gamma}, [ E_{\delta}, E_{-\beta}]] \rangle =0 .
\end{aligned}\end{equation}

From the properties of the Killing form, which follow from the properties of the trace, we obtain $$ R^{\alpha}_{\phantom{*}\gamma, \delta, -\beta}=\langle E_{-\alpha}, [E_{\gamma}, [ E_{\delta}, E_{-\beta}]]\rangle =\langle [ E_{-\alpha}, E_{\gamma}], [ E_{\delta}, E_{-\beta}]\rangle. $$ Then we have
\begin{equation}\label{eq:p-q}\begin{aligned}
& iq^{\alpha}_{xt}- iq^{\alpha}_{xx}  +i q^{\alpha}   +2 q^{\alpha}_x -\frac{1}{a} \sum_{\beta, \gamma, \delta \in \Delta^+_1}
   R^{\alpha}_{\phantom{*}\gamma, \delta, -\beta} q^{\gamma}_x q^{\delta} p^{\beta} =0   , \\
&ip^{\alpha}_{xt}- ip^{\alpha}_{xx}  +i p^{\alpha}  -2 p^{\alpha}_x -\frac{1}{a} \sum_{\beta, \gamma, \delta \in \Delta^+_1}
 R^{-\alpha}_{\phantom{**}-\gamma, \delta, -\beta}   p^{\gamma}_x q^{\delta} p^{\beta} =0 .
\end{aligned}\end{equation}
Variable change can bring the equations to the ``NLS-DNLS-like'' form,
\begin{equation}\label{eq:change}\begin{aligned}
&  q^{\alpha}=\mu e^{-ix} u^{\alpha}  , \\
& p^{\alpha}= \nu e ^{ix} v^{\alpha}.
\end{aligned}\end{equation}
$\mu, \nu$ are some arbitrary (complex) constants. Without loss of generality we can assume $\mu=\nu=1,$ these constants are related to a rescaling of the variables. The equations in terms of $ u^{\alpha} $ and $ v^{\alpha}$ are \begin{equation}\label{eq:u-v}\begin{aligned}
 i u^{\alpha}_{t} +u^{\alpha}_{xx}  - u^{\alpha}_{xt}   -\frac{1 }{a}  \sum_{\beta, \gamma, \delta \in \Delta^+_1}
   R^{\alpha}_{\phantom{*}\gamma, \delta, -\beta} (u^{\gamma}+i u^{\gamma}_x)u^{\delta} v^{\beta} &=0   , \\
-iv^{\alpha}_{t} +v^{\alpha}_{xx}  - v^{\alpha}_{xt} +\frac{1}{a} \sum_{\beta, \gamma, \delta \in \Delta^+_1}
 R^{-\alpha}_{\phantom{**}-\gamma, \delta, -\beta}   (v ^{\gamma}-iv^{\gamma}_x   )u^{\delta} v^{\beta} &=0 .
\end{aligned}\end{equation}

The problem of reductions is an essential one in the theory of integrable systems. The reductions are associated by the action of a finite group of symmetries, known as Mikhailov's reduction group, \cite{Mi}.   Here we point out to the following reductions, noting that there are other possible reductions, such as those discussed in Section \ref{s7}.

R1:
 $p^{\alpha}=\pm \bar{q}^{\alpha}$, giving $v^{\alpha}=\pm \bar{u}^{\alpha}.$ Since ${R} ^{-\alpha}_{\phantom{**}-\gamma, -\delta, \beta}=R^{\alpha}_{\phantom{*}\gamma, \delta, -\beta}$, the equation is

\begin{equation}\label{eq:u-bar}
 iu^{\alpha}_{t}+u^{\alpha}_{xx}  - u^{\alpha}_{xt}   \mp   \frac{1}{a} \sum_{\beta, \gamma, \delta \in \Delta^+_1}
   R^{\alpha}_{\gamma, \delta, -\beta} (u^{\gamma}+iu^{\gamma}_x )u^{\delta} \bar{u}^{\beta} =0    .\end{equation}
When the Cartan-Weyl basis is represented with real matrices, then it is not difficult to check that this equation is $CPT$ - invariant, in a sense that it is invariant under the transformation $u^{\alpha}(x,t) \to \bar{u}^{\alpha}(-x,-t)$.

R2: Reduction leading to a nonlocal equation is possible by taking $v^{\alpha}(x,t)=\pm \bar{u}^{\alpha}(-x,-t) $

\begin{equation}\label{eq:u-nl}
 iu^{\alpha}_{t}(x,t)+u^{\alpha}_{xx}(x,t)  - u^{\alpha}_{xt}(x,t)   \mp   \frac{1}{a} \sum_{\beta, \gamma, \delta \in \Delta^+_1}
   R^{\alpha}_{\phantom{*}\gamma, \delta, -\beta} (u^{\gamma}(x,t)+iu^{\gamma}_x (x,t))u^{\delta}(x,t) \bar{u}^{\beta}(-x,-t) =0    .\end{equation}

The reflections of the time and space variables can be included as elements of the reduction group as well, \cite{V, GeGrIv}.

\section{ Bi-Hamiltonian formulation}\label{s5}

Let us introduce for convenience the variables $m^{\alpha}=u^{\alpha}+iu_x^{\alpha}$ and $n^{\alpha}= v^{\alpha}-iv_x^{\alpha}.$ Then the equations  \eqref{eq:u-v} acquire the form

\begin{equation}\label{eq:m-n}\begin{aligned}
 i m^{\alpha}_{t} +u^{\alpha}_{xx}  -\frac{1 }{a}  \sum_{\beta, \gamma, \delta \in \Delta^+_1}
   R^{\alpha}_{\phantom{*}\gamma, \delta, -\beta}m^{\gamma}u^{\delta} v^{\beta} &=0   , \\
-in^{\alpha}_{t} +v^{\alpha}_{xx}   +\frac{1}{a} \sum_{\beta, \gamma, \delta \in \Delta^+_1}
 R^{-\alpha}_{\phantom{**}-\gamma, \delta, -\beta}   n ^{\gamma}u^{\delta} v^{\beta} &=0 .
\end{aligned}\end{equation}
It is not difficult to check by using \eqref{eq:m-n} that the following quantity is an integral of motion:
\begin{equation} \label{H_1}
\mathcal{H}_1=i \sum_{\alpha \in \Delta_1^+} \int u_x^{\alpha}n^{\alpha}dx = -i \sum_{\alpha \in \Delta_1^+} \int v_x^{\alpha}m^{\alpha}dx.
\end{equation}
One can write $\mathcal{H}_1$ in a covariant form using the metric tensor \eqref{g} and summation convention over the repeated roots from $\Delta_1^+:$

$$\mathcal{H}_1= i \int g _{\alpha, -\beta}u_x^{\alpha}n^{\beta}dx = -i \int g_{-\alpha, \beta}v_x^{\alpha}m^{\beta}dx.$$ We will however use the explicit summation as well when more practical. The integration is over $\mathbb{R}$. Assuming that $u^{\alpha},v^{\alpha}$ decay fast at $|x|\to \infty$ the integration by parts gives
$$ \sum_{\alpha \in \Delta_1^+} \int u_x^{\alpha}n^{\alpha}dx =-\sum_{\alpha \in \Delta_1^+} \int (u^{\alpha}v^{\alpha}_x + i u_x^{\alpha}v_x^{\alpha})dx =- \sum_{\alpha \in \Delta_1^+} \int v_x^{\alpha}m^{\alpha}dx, $$ so the two expressions for $\mathcal{H}_1$ are equal indeed. Furthermore we have
\begin{equation}
\frac{\delta \mathcal{H}_1}{\delta m^{\alpha}}=-iv_x^{\alpha} \qquad \frac{\delta \mathcal{H}_1}{\delta n^{\alpha}}=iu_x^{\alpha}
\end{equation}
The equations \eqref{eq:m-n} could be written in the form
\begin{equation} \label{PB1}
\begin{pmatrix} m_t^{\alpha} \\ n_t ^{\alpha}\end{pmatrix}=\sum_{\beta \in \Delta_1^+}\begin{pmatrix} \mathcal{D}_{11}^{\alpha \beta}  & \mathcal{D}_{12}^{\alpha \beta}\\
\mathcal{D}_{21}^{\alpha \beta} & \mathcal{D}_{22}^{\alpha \beta} \end{pmatrix}\begin{pmatrix} \frac{\delta \mathcal{H}_1}{\delta m^{\beta}} \\ \frac{\delta  \mathcal{H}_1}{\delta n^{\beta}}\end{pmatrix},
\end{equation}
where
\begin{align}
&\mathcal{D}_{11}^{\alpha \beta}=\frac{1 }{a}  \sum_{\gamma, \delta \in \Delta^+_1}
   R^{\alpha}_{\phantom{*}\gamma, \delta, -\beta}m^{\gamma}\partial_x^{-1}m^{\delta}, \nonumber \\
&\mathcal{D}_{12}^{\alpha \beta}=\delta^{\alpha \beta } \partial_x -\frac{1 }{a}  \sum_{\gamma, \delta \in \Delta^+_1}   R^{\alpha}_{\phantom{*}\gamma, \beta, -\delta}m^{\gamma}\partial_x^{-1}n^{\delta} ,\nonumber \\
&\mathcal{D}_{21}^{\alpha \beta}=\delta^{\alpha \beta } \partial_x +\frac{1 }{a}  \sum_{\gamma, \delta \in \Delta^+_1}   R^{-\alpha}_{\phantom{**}-\gamma, \delta, -\beta}n^{\gamma}\partial_x^{-1}m^{\delta} ,\nonumber \\
&\mathcal{D}_{22}^{\alpha \beta}=-\frac{1 }{a}  \sum_{\gamma, \delta \in \Delta^+_1}
   R^{-\alpha}_{\phantom{*}-\gamma, \beta, -\delta}n^{\gamma}\partial_x^{-1}n^{\delta}. \nonumber
\end{align}
Here the Hamiltonian structure $\mathcal{D}^{\alpha \beta}$ coincides with the one from \cite{Ara} and generalizes the Hamiltonian structure $\mathcal{D}$ from \cite{LeFo}. The second Hamiltonian structure is proportional to the Hamiltonian operator $\mathcal{E}$ from \cite{LeFo}

\begin{equation}\label{op_E}
\begin{pmatrix} m_t^{\alpha} \\ n_t ^{\alpha}\end{pmatrix}=\sum_{\beta \in \Delta_1^+}\begin{pmatrix} 0  & -i(1+i \partial_x)\delta^{\alpha \beta}\\
i(1-i \partial_x)\delta^{\alpha \beta} & 0 \end{pmatrix}\begin{pmatrix} \frac{\delta \mathcal{H}_2}{\delta m^{\beta}} \\ \frac{\delta \mathcal{H}_2}{\delta n^{\beta}}\end{pmatrix},
\end{equation}
The above representation is equivalent to
\begin{equation}\label{op_E1}
 m_t^{\alpha} =-i\frac{\delta \mathcal{H}_2}{\delta v^{\alpha}}, \qquad
 n_t ^{\alpha} = i \frac{\delta \mathcal{H}_2}{\delta u^{\alpha}} ,
\end{equation}
where the functional $\mathcal{H}_2$ is the second Hamiltonian
\begin{align} \label{H_2}
\mathcal{H}_2 &= \sum_{\alpha \in \Delta_1^+} \int \left( - u_{xx}^{\alpha}v^{\alpha} +\frac{1 }{2a}  \sum_{\beta, \gamma, \delta \in \Delta^+_1}
   R_{-\alpha, \gamma, \delta, -\beta} m^{\gamma}u^{\delta} v^{\beta}v^{\alpha} \right) dx \nonumber \\
& = \sum_{\alpha \in \Delta_1^+} \int \left( - u^{\alpha}v_{xx}^{\alpha} +\frac{1 }{2a}  \sum_{\beta, \gamma, \delta \in \Delta^+_1}
   R_{\alpha, -\gamma, -\beta, \delta } n^{\gamma} v^{\beta} u^{\delta}u^{\alpha} \right) dx.
\end{align}
One can check that the two expressions are equal by using the fact that the following integral is zero:
\begin{equation}
I=\sum_{\alpha ,\beta, \gamma, \delta \in \Delta^+_1}
   \int R_{-\alpha, \gamma, \delta, -\beta} (v^{\alpha}u^{\gamma})_x v^{\beta}u^{\delta}  dx .
\end{equation}
With integration by parts and relabeling of the summation indices, using also the properties of the quantities $R_{*, *, *, *}$ one can verify that $I=-I$ which implies $I=0$.

The functional derivatives \eqref{op_E1} contain two terms with $R_{*, *, *, *}$ and the property \eqref{pr2} plus relabeling can be used to demonstrate that they are equal.

Another conserved quantity which can be checked directly by \eqref{eq:m-n} is \begin{equation} \label{H_0}
\mathcal{H}_0= \sum_{\alpha \in \Delta_1^+} \int m^{\alpha}n^{\alpha}dx .
\end{equation}
We point out that throughout the reduction procedures R1 and R2 the Hamiltonian structures and the Hamiltonians do not degenerate.

\section{Examples of multi-component Fokas-Lenells equations} \label{s6}

We describe in details the multi-component Fokas-Lenells equations related to each of the four classes of Hermitian symmetric spaces:  A.III, BD.I, C.I and D.III.

\subsection{A.III, $SU(n+1)/S(U(1)\times U(n))$  }
In this and next subsections we will consider $SU(n+m)/S(U(m)\otimes U(n))$. It is well known that the set of positive roots of the algebra $su(n+m)$ is given by $\Delta^+ \equiv \{ e_j - e_k , 1 \leq j < k \leq n+m\}$ \cite{h}.
We start with the special case $m=1.$  The root subsystems are:
\begin{equation}\label{eq:A31}\begin{split}
 \Delta_0^+ = \{ e_k - e_j, \quad 2 \leq  k < j \leq n+1\}, \qquad \Delta_1^+ = \{ e_1 - e_j, \quad 2 \leq j \leq n+1\}.
\end{split}\end{equation}
where $e_i$ are orthonormal basis vectors in $n+1$ dimensional Euclidean space with the usual scalar product.

Let us choose the special element  $J\in \mathfrak{h}$ as dual to the vector  $$\vec{J}= \frac{1}{n+1}(ne_1-e_2-e_3-\ldots-e_{n+1}).$$
It is easy to see that all roots $\alpha \in \Delta_0^+$ satisfy $(\alpha,\vec{J})=0$, while all roots $\beta \in \Delta_1^+$ satisfy $(\beta,\vec{J})=1$, hence $a=1.$

The corresponding Cartan-Weyl generators are (with the convention $(E_{i,j})_{i',j'} = \delta_{i,i'}\delta_{j,j'} $ where $\delta_{ij}$ is the Kronecker delta)
\begin{equation} \label{CW}
\begin{split}
E_{\alpha} \equiv E_{e_k - e_j}=E_{k,j}, \qquad E_{-\alpha}=E_{j,k}, \qquad H_{\alpha} = E_{kk} - E_{jj}.
\end{split}
\end{equation}
The Killing form is given by $ \langle X,Y \rangle = \text{tr}(XY) $.
Assuming now $\alpha_k=e_1-e_{k+1}$ we evaluate $$[E_{-\alpha_k}, E_{\alpha_l}]=E_{k+1,l+1}-E_{11}\delta_{kl}  $$ which implies
$$R^{\alpha_k}_{\phantom{**}\alpha_l ,\alpha_r, -\alpha_s}=-(\delta_{lk}\delta_{rs}+\delta_{kr}\delta_{ls}), \qquad R^{-\alpha_k}_{\phantom{**}-\alpha_l ,\alpha_r, -\alpha_s}=\delta_{lk}\delta_{rs}+\delta_{lr}\delta_{ks}.$$
We adopt further the notation $q^{\alpha_k} \equiv q^k, $  $p^{\alpha_k} \equiv p^k, $ $u^{\alpha_k} \equiv u^k, $ $v^{\alpha_k} \equiv v^k, $  $k=1, \ldots, n.$ Hence, one can consider a vector notations $\vec{q}=(q^1, \ldots, q^n)^T  $  etc. The matrices $J, Q $ are $(n+1)\times(n+1)$ dimensional and have the following obvious block-structure, the lower-right block is $n\times n$ dimensional:
\begin{equation}\label{eq:QJ}\begin{split}
Q = \left(\begin{array}{cc} 0 & \vec{q} ^T \\ \vec{p}  & 0   \end{array}\right) , \qquad
  J = \frac{1}{n+1}\left(\begin{array}{cc} n & 0 \\ 0 & -\openone   \end{array}\right).
\end{split}\end{equation}
From \eqref{eq:p-q} we obtain the equations
\begin{equation}\label{eq:p-q1}\begin{aligned}
& i(\vec{q}_{xt}- \vec{q}_{xx} - \vec{ q})   +2 \vec{q}_x +\big((\vec{p}\cdot \vec{q}) \vec{q}_x + (\vec{p}\cdot \vec{q}_x) \vec{q}\big ) =0   , \\
&i(\vec{p}_{xt}- \vec{p}_{xx} - \vec{ p})   -2 \vec{p}_x -\big((\vec{p}\cdot \vec{q}) \vec{p}_x + (\vec{p}_x \cdot \vec{q}) \vec{p}\big ) =0  .
\end{aligned}\end{equation}
which can be written in terms of $\vec{u}$ and $\vec{v}$ according to \eqref{eq:u-v}: \begin{equation}\label{eq:u-v1}\begin{aligned}
 i\vec{u}_{t}+ \vec{u}_{xx} - \vec{ u}_{xt} +\big(2 (\vec{u}\cdot \vec{v}) \vec{u}+(\vec{u}\cdot \vec{v})i \vec{u}_x +i(\vec{u}_x\cdot \vec{v}) \vec{u} \big ) =0&   , \\
  -i\vec{v}_{t}+ \vec{v}_{xx} - \vec{ v}_{xt} +\big(2 (\vec{u}\cdot \vec{v}) \vec{v}-(\vec{u}\cdot \vec{v})i \vec{v}_x -i(\vec{u}\cdot \vec{v}_x) \vec{v} \big ) =0 &  .
\end{aligned}\end{equation}

Introducing $\vec{m}=\vec{u}+i\vec{u}_x$, $\vec{n}=\vec{v}-i\vec{v}_x$, the first Hamiltonian is

\begin{equation} \label{H11}
\mathcal{H}_1=i  \int \vec{u}_x \cdot \vec{n} \, dx = -i  \int \vec{v}_x \cdot \vec{m} \,dx,
\end{equation}  the first Hamiltonian structure is
\begin{equation}
\begin{pmatrix} m_t^{k} \\ n_t ^{k}\end{pmatrix}=\sum_{s=1}^{n}\begin{pmatrix} -(m^k \partial_x ^{-1} m^s +m^s \partial_x ^{-1} m^k)  & (\partial_x +  \vec{ m} \cdot  \partial_x ^{-1} \vec{ n})\delta^{ks}+m^k \partial_x ^{-1} n^s\\
(\partial_x +  \vec{ m} \cdot  \partial_x ^{-1} \vec{ n})\delta^{ks}+ n^k \partial_x ^{-1} m^s & -(n^k \partial_x ^{-1} n^s +n^s \partial_x ^{-1} n^k) \end{pmatrix}\begin{pmatrix} \frac{\delta \mathcal{H}_1}{\delta m^{s}} \\ \frac{\delta \mathcal{H}_1}{\delta n^{s}}\end{pmatrix},
\nonumber \end{equation} and coincides with the one from \cite{Ara}. The second Hamiltonian is
$$\mathcal{H}_2= -\int \big(\vec{u}_{xx}\cdot \vec{v} + (\vec{m}\cdot \vec{v})(\vec{u}\cdot \vec{v}) \big)dx=-\int \big(\vec{u}\cdot \vec{v}_{xx} + (\vec{n}\cdot \vec{u})(\vec{u}\cdot \vec{v}) \big)dx,$$
the corresponding Hamiltonian structure is obvious from \eqref{op_E} after replacement of the indices $\alpha \to k,$ $\beta \to s.$

The two reductions of \eqref{eq:u-v1} are:

R1. Using Hermitian conjugation ($\dagger$) and matrix notations for the scalar products and bold face for the (complex-valued) vector-columns we have  \eqref{eq:u-bar} in the form
\begin{equation}\label{eq:u-R1}
 i{\bf u}_{t}+ {\bf u}_{xx} - {\bf u}_{xt} \pm  \big(2 ({\bf u}^{\dagger} {\bf u}) {\bf u}+({\bf u}^{\dagger} {\bf u})i {\bf u}_x +i({\bf u}^{\dagger} {\bf u}_x ) {\bf u} \big ) =0.
\end{equation}
This equation is equivalent to (82) of \cite{Guo} after a change of variables, indicated in \cite{Guo}.

R2: Equation \eqref{eq:u-nl} in this case is
\begin{equation}\label{eq:u-R2}
 i{\bf u}_{t}+ {\bf u}_{xx} - {\bf u}_{xt} \pm \big(2 (\tilde{\bf u}^{T} {\bf u}) {\bf u}+(\tilde{\bf u}^{T} {\bf u})i {\bf u}_x +i(\tilde{\bf u}^{T} {\bf u}_x ) {\bf u} \big ) =0
\end{equation}
where ${\bf u}={\bf u}(x,t)$  and $\tilde{\bf u}=\bar{\bf u}(-x,-t).$

\subsection{ A.III symmetric spaces $SU(m+n)/S(U(m)\times U(n))$ }\label{A1}

In order to describe the local coordinates of this symmetric space $\mathfrak{g}^{(1)},$ cf. \eqref{eq:g0}, we need the structure of the algebra $su(m+n)$ and its subalgebras $su(m) \oplus su(n)$.
In what follows we split the set of indices $\mathcal{K}\equiv \{ 1, 2, \dots, m+n\}$ into two subsets: $\mathcal{K}=\mathcal{K}_1\cup \mathcal{K}_2$
where $\mathcal{K}_1\equiv \{ 1, 2, \dots, m\}$ and $\mathcal{K}_2\equiv \{m+1, m+2, \dots, m+n\}$. Also we denote by $a, b, c$ the indices with values in $\mathcal{K}_1$,
by $j,k,l$ - the indices taking values in $\mathcal{K}_2$ and the indices $s,v$  take values in $\mathcal{K}$.

With these notation the set of positive roots of $su(m+n)$, $su(m)\oplus su(n)$ is:
\begin{equation}\label{eq:Delp}\begin{split}
\Delta^+_0 \equiv \{ e_a - e_b, a< b\} \cup  \{ e_j - e_k, j< k\}, \qquad  \Delta_1^+ \equiv \{ e_a - e_j \}.
\end{split}\end{equation}
In fact, the algebra $\mathfrak{g}\equiv su(m+n)$ acquires $\mathbb{Z}_2$ grading: $\mathfrak{g} \equiv \mathfrak{g}^{(0)}\oplus \mathfrak{g}^{(1)}$,
where $\mathfrak{g}^{(0)}=s(u(m) \oplus u(n))$ and the linear space $\mathfrak{g}^{(1)}$ will be described below. The Cartan-Weyl basis of $\mathfrak{g}$ is given in eq. (\ref{CW}).
This grading is directly related to the Cartan subalgebra element $J_0$:
\begin{equation}\label{eq:J}\begin{split}
 J_0 = \frac{1}{m+n} \left( n\sum_{a=1}^{m} E_{a,a} - m \sum_{k=m+1}^{m+n} E_{k,k} \right) = \frac{1}{m+n} \left(\begin{array}{cc} n \openone_m & 0 \\ 0 & -m\openone_m \end{array}\right).
\end{split}\end{equation}
$J_0$ is dual to the vector $\vec{J}_0 $
\begin{equation}\label{eq:j}\begin{split}
\vec{J}_0 = \frac{1}{m+n} \left( n\sum_{a=1}^{m} e_{a} - m \sum_{k=m+1}^{m+n} e_{k} \right).
\end{split}\end{equation}
Therefore
\begin{equation}\label{eq:A301}\begin{split}
 \alpha (J_0) = (\vec{J}_0,\alpha) =0, \quad \alpha \in \Delta_0, \qquad  \beta (J_0) = (\vec{J}_0,\beta) =1, \quad \beta \in \Delta_1^{+},\quad a=1.
\end{split}\end{equation}

We are using notations in which the typical representation of $su(m+n)$ is a set of $(m+n)\times (m+n)$ matrices with an obvious block-matrix
 structure: $\mathfrak{g}^{(0)} \equiv s(u(m)\bigoplus u(n))$ i.e. $\mathfrak{g}^{(0)}$ consists of traceless block-diagonal matrices, while the linear space $\mathfrak{g}^{(1)}$ is spanned by block-off-diagonal matrices:
 \begin{equation}\label{eq:Qtz}\begin{split}
\mathfrak{g}^{(0)} \simeq \left(\begin{array}{cc} u(m) & 0\\ 0 & u(n)   \end{array}\right), \qquad \mathfrak{g}^{(1)} \simeq \left(\begin{array}{cc} 0  &  \q \\ \p & 0   \end{array}\right).
\end{split}\end{equation}

With  $J$ and $ Q $ taken as matrices in the form
\begin{equation}\label{eq:A3mJQ}\begin{split}
 J =J_0= \frac{1}{m+n} \left(\begin{array}{cc} n \openone_m & 0 \\ 0 & -m\openone_m \end{array}\right) \in \mathfrak{g}^{(0)}, \quad   Q(x,t) = \left(\begin{array}{cc} 0 &  \q \\  \p  &  0  \end{array}\right) \in \mathfrak{g}^{(1)}
\end{split}\end{equation} following \eqref{eq:gen} we obtain
\begin{equation}\label{eq:SpV}\begin{split}
V_{-1}(x,t) = i \left(\begin{array}{cc} 0 &  \q \\  -\p  &  0  \end{array}\right), \quad
 V_0(x,t) =\left(\begin{array}{ccc} \q \p & 0 \\ 0 & -\p \q   \end{array}\right),
\end{split}\end{equation}
as well as the equations in block-matrix form
\begin{equation}\label{eq:p-q_A3m}\begin{aligned}
& i\left( \q_{xt}- \q_{xx} +\q \right)   + 2 \q_x + (\q_{x} \p \q + \q \p \q_{x} ) =0  , \\
& i\left( \p_{xt}- \p_{xx} + \p \right) -2\p_x  - (\p_{x} \q \p + \p \q \p_{x})  =0.
\end{aligned}\end{equation}
Introducing new matrices $\u,\v$ such that
$$\q=e^{-ix}\u, \qquad \p=e^{ix}\v $$ we represent the equations \eqref{eq:p-q_A3m}
in the form
\begin{equation}\label{eq:u-v_A3m}\begin{aligned}
 i\u_t -\u_{xt}+ \u_{xx}  + (\u+i\u_{x})\v\u+ \u\v (\u+i\u_x)=&0  , \\
 -i \v_t - \v_{xt}+ \v_{xx} +(\v-i\v_{x}) \u \v + \v \u (\v-i\v_{x})  =&0.
\end{aligned}\end{equation}
For $\u$ and $\v,$ and $\m=\u+i\u_x$ and $\n=\v-i\v_x,$ there is natural embedding such that\footnote{The use of the letter $U$ here should not be confused with its use in the other contexts.}
\begin{equation}\label{eq:SpV'}\begin{split}
&U =  \left(\begin{array}{cc} 0 &  \u \\  0  &  0  \end{array}\right)=\sum_{\alpha \in \Delta_1^+ } U^{\alpha}E_{\alpha}, \quad
 M =\left(\begin{array}{ccc} 0 & \m \\ 0 &  0   \end{array}\right)=\sum_{\alpha \in \Delta_1^+ } M^{\alpha}E_{\alpha} ,  \\
&V =  \left(\begin{array}{cc} 0 &  0 \\  \v  &  0  \end{array}\right)=\sum_{\alpha \in \Delta_1^+ } V^{\alpha}E_{-\alpha}, \quad
 N =\left(\begin{array}{ccc} 0 & 0 \\ \n &  0   \end{array}\right)=\sum_{\alpha \in \Delta_1^+ } N^{\alpha}E_{-\alpha} , \quad \text{etc.}
\end{split}\end{equation}

The Hamiltonians are as follows

\begin{equation} \label{C-H_1}
\mathcal{H}_1=i \sum_{\alpha \in \Delta_1^+} \int U_x^{\alpha}N^{\alpha}dx = i\int \langle U_x N \rangle dx =i \int \mathrm{tr}(\u_x \n) dx
\end{equation} and similarly
\begin{equation}
\mathcal{H}_1=- i\int \langle V_x M \rangle dx=- i \int \mathrm{tr}(\v_x \m)dx;
\end{equation}

\begin{align} \label{C-H_2}
\mathcal{H}_2 &= \sum_{\alpha \in \Delta_1^+} \int \left( - U_{xx}^{\alpha}V^{\alpha} +\frac{1 }{2}  \sum_{\beta, \gamma, \delta \in \Delta^+_1}
   R_{-\alpha, \gamma, \delta, -\beta} M^{\gamma}U^{\delta} V^{\beta}V^{\alpha} \right) dx \nonumber \\
&=  \int \left( -\langle U_{xx},V \rangle +\frac{1 }{2}  \sum_{\alpha , \beta, \gamma, \delta \in \Delta^+_1}
   \langle [ E_{-\alpha}, E_{\gamma}], [ E_{\delta}, E_{-\beta}]\rangle M^{\gamma}U^{\delta} V^{\beta}V^{\alpha} \right) dx \nonumber \\
&=  \int \left( -\langle U_{xx},V \rangle +\frac{1 }{2} \langle [ V, M], [ U, V]\rangle  \right) dx \nonumber \\
&=  -\int \left( \mathrm{tr}( \u_{xx}\v ) + \mathrm{tr}( \m \v \u \v ) \right) dx
\end{align} and
\begin{align} \label{C-H_2'}
\mathcal{H}_2&= \int \left( -\langle U,V_{xx} \rangle +\frac{1 }{2} \langle [U, N], [ V, U]\rangle  \right) dx= -\int \left( \mathrm{tr}( \u\v_{xx} ) +\mathrm{tr}( \n \u \v \u  ) \right) dx.
\end{align}

\begin{remark}\label{rem:2} For a functional  $ f=\int \langle \a, \b \rangle dx=\int \mathrm{tr}( \a\b )dx=\sum_{i,j} \int \a_{ij}\b_{ji}dx $ we have $$\a_{ij}=\frac{\delta f}{\delta \b_{ji}} \text{ and hence we write } \a= \frac{\delta f}{\delta \b^T}. $$
\end{remark}

Following \eqref{PB1} one can represent the equations \eqref{eq:u-v_A3m} in the form
\begin{equation}\label{eq:M-N_A3m}\begin{aligned}
& M_t =\left(\frac{\delta \mathcal{H}_1}{\delta N^T}\right)_x  +\left[ M, \partial_x^{-1}\left[ M,  \frac{\delta \mathcal{H}_1}{\delta M^T}\right]
+\partial_x^{-1}\left[ N,  \frac{\delta \mathcal{H}_1}{\delta N^T}\right]   \right] , \\
 &N_t =\left(\frac{\delta \mathcal{H}_1}{\delta M^T}\right)_x  +\left[ N, \partial_x^{-1}\left[ M,  \frac{\delta \mathcal{H}_1}{\delta M^T}\right]
+\partial_x^{-1}\left[ N,  \frac{\delta \mathcal{H}_1}{\delta N^T}\right]   \right] .
\end{aligned}\end{equation}
This leads to the following representation of the first Hamiltonian structure \eqref{PB1} in the matrix case:

\begin{equation} \label{PB1m}
\begin{pmatrix} M_t \\ N_t \end{pmatrix}=\begin{pmatrix} \mathrm{ad}_M \partial_x^{-1}\mathrm{ad}_M  \quad & \partial_x + \mathrm{ad}_M \partial_x^{-1}\mathrm{ad}_N\\
\partial_x + \mathrm{ad}_N \partial_x^{-1}\mathrm{ad}_M &  \mathrm{ad}_N \partial_x^{-1}\mathrm{ad}_N \end{pmatrix}\begin{pmatrix} \frac{\delta \mathcal{H}_1}{\delta M^T} \\ \frac{\delta  \mathcal{H}_1}{\delta N^T}\end{pmatrix}.
\end{equation}  The equations \eqref{eq:u-v_A3m} could be written as
\begin{equation}\label{op_E1m}
 \m_t =-i\frac{\delta \mathcal{H}_2}{\delta \v^{T}}, \qquad
 \n_t  = i \frac{\delta \mathcal{H}_2}{\delta \u^{T}} ,
\end{equation}which allows to represent the second Hamiltonian structure from \eqref{op_E} as
\begin{equation}\label{op_Em}
\begin{pmatrix} M_t \\ N_t \end{pmatrix}=\begin{pmatrix} 0  & -i(1+i \partial_x)\\
i(1-i \partial_x) & 0 \end{pmatrix}\begin{pmatrix} \frac{\delta \mathcal{H}_2}{\delta M^{T}} \\ \frac{\delta \mathcal{H}_2}{\delta N^{T}}\end{pmatrix}.
\end{equation}

The first reduction involves Hermitian conjugation $\v=\pm \u^{\dagger},$  the equations \eqref{eq:u-v_A3m}  reduce to
 \begin{equation}\label{eq:R1_A3}
 i\u_t -\u_{xt}+ \u_{xx}  \pm (2\u\u^{\dagger}\u+i\u_{x}\u^{\dagger}\u + i\u\u^{\dagger} \u_x)=0.
\end{equation}
The second reduction $\v(x,t)=\pm \u^{\dagger}(-x,-t),$ leads to the following nonlocal equation
\begin{equation}\label{eq:R2_A3}
 i\u_t -\u_{xt}+ \u_{xx}  \pm (2\u \tilde{\u} \u +i\u_{x}\tilde{\u}\u+ i\u \tilde{\u}\u_x)=0,
  \end{equation}
where $\tilde{\u}={\u}^{\dagger}(-x,-t).$

\subsection{BD.I symmetric space $SO(2n+1)/(SO(2n-1)\times SO(2))$}

We are using the following matrix realization of the typical representation of the Lie algebra $so(2n+1).$
Introducing the $(2n+1)\times(2n+1)$ matrix
\begin{equation}\label{S}
S = \sum_{k=1}^{2n+1}(-1)^{k+1}E_{k\bar{k}} =\left(\begin{array}{ccc}0 & 0 & 1 \\ 0 & -s_0 & 0 \\ 1& 0 & 0 \end{array}\right), \qquad \bar{k}=2n+2-k , \qquad S^{-1}=S,
\end{equation} the $so(2n+1)$ algebra is the set of all $(2n+1)\times(2n+1)$ matrices $X$, which satisfy
$$X+ SX^TS^{-1}=0.$$
With this definition  the generators of the Cartan subalgebra $H_{e_p}$ are diagonal, see eq. (\ref{eq:W_BD}) below.


The element, that corresponds to $J$ in the root space is $\vec{J}=e_1,$ (and $J=H_{e_1}).$ The
corresponding subsets $\Delta_0^+$ and $\Delta_1^+$ of the root system are
\begin{equation}\label{eq:BD11}\begin{split}
 \Delta_0^+ = \{ e_k - e_j,\; e_k+e_j ,\; 2e_k \quad 2 \leq  k < j \leq n\}, \qquad \Delta_1^+ = \{ e_1 - e_k,\; e_1+e_k, \; e_1 \quad 2 \leq k \leq n\}.
\end{split}\end{equation}
All roots $\alpha \in \Delta_0^+$ satisfy $(\alpha,\vec{J})=0$, while all roots $\beta \in \Delta_1^+$ satisfy $(\beta,\vec{J})=1,$ therefore in this case $a=1.$
 Now $\Delta_1^+$ contains $2n-1$ positive roots. The Cartan-Weyl generators can be represented in the form
\begin{equation}\label{eq:W_BD}\begin{aligned}
E_{e_k-e_j} &= E_{kj} - (-1)^{k+j} E_{\bar{j},\bar{k}}, &\quad E_{e_k+e_j} &= E_{k,\bar{j}} - (-1)^{k+j} E_{j,\bar{k}}, \\
E_{e_k} &= E_{k,n+1} +(-1)^{n+k} E_{n+1,\bar{k}}, &\quad H_{e_k} &= E_{k,k} - E_{\bar{k},\bar{k}},
\end{aligned}\end{equation}
and in addition $E_{-\alpha} = E_\alpha^T$.
The Killing form is given by $$ \langle X,Y \rangle = \frac{1}{2}\text{tr}(XY) $$ and on the Cartan-Weyl generators in their representation \eqref{eq:W_BD} satisfies \eqref{eq:KF}.

The vectors of the root subspace $\Delta_1^+$ could be labeled as follows: ${\alpha}_k=e_1 - e_{k},$  $k=2,3,\ldots, n$ (i.e. $n-1$ roots of this type); ${\alpha}_{n+1}=e_1$ and ${\alpha}_{\bar{k}} =e_1+ e_{k},$  $k=2,3,\ldots,n$ and $\bar{k}=2n+2-k$ as before. Note that $\overline{n+1}=n+1,$ i.e. there are $2n-1$ positive roots.   Moreover $a=(\vec{J},{\alpha}_k)=1$ for all $k=2,\ldots,2n     .$
We evaluate $$[E_{-\alpha_k}, E_{\alpha_l}]=E_{kl}-(-1)^{k+l}E_{\bar{l}, \bar{k}} +(E_{\bar{1} \bar{1}}-E_{11})\delta_{kl} $$ allowing us to obtain
$$R^{\alpha_k}_{\phantom{**}\alpha_l ,\alpha_m, -\alpha_s}=(-1)^{m+s}\delta_{k\bar{s}}\delta_{l\bar{m}}-\delta_{km}\delta_{ls}
-\delta_{kl}\delta_{sm}$$ and similarly
$$R^{-\alpha_k}_{\phantom{**}-\alpha_l ,\alpha_m, -\alpha_s}=\delta_{kl}\delta_{sm}+\delta_{lm}\delta_{ks}
-(-1)^{k+l}\delta_{\bar{k}m}\delta_{\bar{l}s}.$$
We adopt the notation: $q^{\alpha_k} \equiv q^k, $  $p^{\alpha_k} \equiv p^k, $ $u^{\alpha_k} \equiv u^k, $ $v^{\alpha_k} \equiv v^k, $  $k=2, \ldots, 2n.$ Hence, one can introduce  vector notations $\vec{q}=(q^2, \ldots, q^{2n})^T , $  a $2n-1$ component vector etc.

The matrices $J$ and $Q $ are $(2n+1)\times (2n+1)$  matrices with the following block structure:
\begin{equation}\label{eq:BD1JQ}\begin{split}
Q(x,t) = \left(\begin{array}{ccc} 0 & \vec{q}^T & 0 \\  \vec{p} & 0 & s_0 \vec{q} \\ 0 & \vec{p}^T s_0 & 0 \end{array}\right), \quad
 J = \left(\begin{array}{ccc} 1 & 0 & 0 \\ 0 & 0 & 0 \\ 0 & 0 & -1   \end{array}\right).
\end{split}\end{equation}
where the central block of zeroes has dimensionality $(2n-1)\times(2n-1)$ and
$s_0$ is  $(2n-1)\times(2n-1)$ matrix
\begin{equation}\label{S0}
s_0 = \sum_{k=1}^{2n-1}(-1)^{k}E_{k\bar{k}}, \qquad \bar{k}=2n+2-k .
\end{equation}

The equations arising from \eqref{eq:p-q} are
\begin{equation}\label{eq:p-q_BD}\begin{aligned}
& i(\vec{q}_{xt}- \vec{q}_{xx} - \vec{ q})   +2 \vec{q}_x +\big((\vec{p}\cdot \vec{q}) \vec{q}_x + (\vec{p}\cdot \vec{q}_x) \vec{q} -(\vec{q}\cdot s_0\vec{q}_x )s_0\vec{p} \big ) =0   , \\
&i(\vec{p}_{xt}- \vec{p}_{xx} - \vec{ p})   -2 \vec{p}_x -\big((\vec{p}\cdot \vec{q}) \vec{p}_x + (\vec{p}_x \cdot \vec{q}) \vec{p}-(\vec{p}\cdot s_0\vec{p}_x )s_0\vec{q}\big ) =0
\end{aligned}\end{equation}
where the scalar product is $\vec{p}\cdot \vec{q}=\sum _{k=2}^{2n}p^k q^k $.

The equations \eqref{eq:u-v} in this case are
\begin{equation}\label{eq:u-v_BD1}\begin{aligned}
 i\vec{u}_{t}+ \vec{u}_{xx} - \vec{ u}_{xt} +\big((\vec{u}\cdot \vec{v})( \vec{u}+i \vec{u}_x)  +(\vec{v}\cdot \vec{u})\vec{u} + i(\vec{v}\cdot \vec{u}_x)\vec{u} -(\vec{u}\cdot s_0\vec{u} )s_0\vec{v} -i(\vec{u}\cdot s_0\vec{u}_x) s_0\vec{v} \big ) =0&   , \\
  -i\vec{v}_{t}+ \vec{v}_{xx} - \vec{ v}_{xt} +\big((\vec{v}\cdot \vec{u})( \vec{v}-i \vec{v}_x)  +(\vec{u}\cdot \vec{v})\vec{v} - i(\vec{u}\cdot \vec{v}_x)\vec{v} -(\vec{v}\cdot s_0\vec{v} )s_0\vec{u} +i(\vec{v}\cdot s_0\vec{v}_x) s_0\vec{u} \big ) =0 &  .
\end{aligned}\end{equation}
In terms of $\vec{m}=\vec{u}+i\vec{u}_x$, $\vec{n}=\vec{v}-i\vec{v}_x$, the first and the second Hamiltonians are
\begin{align} \label{H_bd}
\mathcal{H}_1&=i  \int \vec{u}_x \cdot \vec{n} \, dx = -i  \int \vec{v}_x \cdot \vec{m} \,dx, \\
\mathcal{H}_2&= -\int \left(\vec{u}_{xx}\cdot \vec{v} + \frac{1}{2} \big(2(\vec{m}\cdot \vec{v})(\vec{u}\cdot \vec{v})-(\vec{v} \cdot s_0 \vec{v})(\vec{m} \cdot s_0 \vec{u})\big)\right)dx\\
&=-\int \left(\vec{u}\cdot \vec{v}_{xx} + \frac{1}{2} \big(2(\vec{n}\cdot \vec{u})(\vec{u}\cdot \vec{v}) - (\vec{u} \cdot s_0 \vec{u})(\vec{n} \cdot s_0 \vec{v})\big)\right)dx .
\end{align}

The reductions in this case are:

R1. The reduction \eqref{eq:u-bar} for the complex-valued vector-column ${\bf u}$ gives
\begin{equation}\label{eq:R1_BD1}
 i{\bf u}_{t}+ {\bf u}_{xx} - {\bf u}_{xt} \pm \big(({\bf u}^{\dagger} {\bf u})i {\bf u}_x  +2({\bf u}^{\dagger} {\bf u} ){\bf u} + i({\bf u}^{\dagger} {\bf u}_x ){\bf u} -({\bf u}^{T} s_0{\bf u} )s_0 \bar{\bf u} -i({\bf u}^T s_0{\bf u} _x) s_0 \bar{\bf u} \big ) =0
  \end{equation}

R2: The second reduction \eqref{eq:u-nl} leads to the nonlocal version of the equation
\begin{equation}\label{eq:R1_BD1nl}
 i{\bf u}_{t}+ {\bf u}_{xx} - {\bf u}_{xt} \pm \big((\tilde{\bf u}^{T} {\bf u})i {\bf u}_x  +2(\tilde{\bf u}^{T} {\bf u} ){\bf u} + i(\tilde{\bf u}^{T} {\bf u}_x ){\bf u} -({\bf u}^{T} s_0{\bf u} )s_0 \tilde{{\bf u}} -i({\bf u}^T s_0{\bf u} _x) s_0 \tilde{{\bf u}} \big ) =0
  \end{equation}
where ${\bf u}={\bf u}(x,t)$  and $\tilde{\bf u}=\bar {\bf u}(-x,-t).$

\subsection{C.I symmetric space $SP(2n)/SU(n)$}\label{c1}

{The $sp(2n)$ algebra may be represented as the set of all $2n\times 2n$ matrices $Y$, which satisfy $$Y+ S_1Y^TS_1^{-1}=0,$$ where $S_1$ is the $2n\times 2n$ matrix
\begin{equation}\label{Sc}
S_1 = \sum_{k=1}^{2n}(-1)^{k+1}E_{k\bar{k}}, \qquad \bar{k}=2n+1-k , \qquad S_1^{-1}=-S_1.
\end{equation}
The $S_1$ matrix possesses a square-block structure
\begin{equation}\label{eq:S1}
S_1 = \left(\begin{array}{cc} 0 &  \s_{12} \\  \s_{21}  &  0  \end{array}\right),\qquad s_{21}=-s_{12}^{-1} =(-1)^n s_{12}\end{equation} where $\s_{ij}$ are square $n\times n$ matrices.

The choice of Cartan involution in this case is related to $\vec{J}=\frac{1}{2} \sum_{s=1}^{n}{e_s}.$
The corresponding subsets $\Delta_0^+$ and $\Delta_1^+$ of the root system are
\begin{equation}\label{eq:C11}\begin{split}
 \Delta_0^+ = \{ e_k - e_j,\quad 1 \leq  k < j \leq n\}, \qquad \Delta_1^+ = \{ e_k + e_j,\;\quad 1 \leq  k < j \leq n, \quad  2e_k, \quad 1 \leq k \leq n\}.
\end{split}\end{equation}
Thus $\Delta_1^+$ contains $n(n+1)/2$ positive roots. The roots $\alpha \in \Delta_0^+$ satisfy $(\alpha,\vec{J})=0$, while all roots $\beta \in \Delta_1^+$ satisfy $(\beta,\vec{J})=1,$ therefore $a=1.$ The Cartan-Weyl generators can be represented in the form
\begin{equation}\label{eq:CW_CC}\begin{aligned}
E_{e_k-e_j} &= E_{kj} - (-1)^{k+j} E_{\bar{j},\bar{k}}, &\quad E_{e_k+e_j} &= E_{k,\bar{j}} + (-1)^{k+j} E_{j,\bar{k}}, \\
E_{2e_k} &= \sqrt{2} E_{k,\bar{k}} , &\quad H_{e_k} &= E_{k,k} - E_{\bar{k},\bar{k}},
\end{aligned}\end{equation}
where $\bar{k} = 2k+1-k$ and in addition $E_{-\alpha} = E_\alpha^T$.
The Killing form is given by \begin{equation} \label{KillingC} \langle X,Y \rangle = \frac{1}{2}\text{tr}(XY). \end{equation}

A general element
\begin{equation} \label{expansion-gen} Q=\sum_{\alpha \in \Delta_1^+ } \langle Q, E_{-\alpha}\rangle E_{\alpha}+\langle Q, E_{\alpha}\rangle E_{-\alpha}= \left(\begin{array}{cc} 0 &  \q \\  \p  &  0  \end{array}\right),
\end{equation} of the symmetric space has the following block-off-diagonal form with $\p,\q$ being square $n\times n$ matrices satisfying
 \begin{equation}\label{eq:sp-pq}
\q+\s_{12}\q^T \s_{12}^{-1}=0, \qquad \p+\s_{21}\p^T
\s_{21}^{-1}=0.\end{equation}
In other words, the components of the matrices $\p$ and $\q$ are not independent, they have $n(n+1)/2$ independent components, one for each root vector $\alpha\in \Delta_1^+$.

For $n=2$ for example, each of the blocks $\q$ and $\p$ is parametrized by 3 matrix elements  as follows (we introduce the notation $\underline{k}=n-k+1$):
\begin{equation}\label{eq:qp2}\begin{split}
\q = \left(\begin{array}{ccc}  q_{1\underline{2}} & \sqrt{2} q_{1\underline{1}} \\ \sqrt{2} q_{2\underline{2}} & -q_{1\underline{2}} \end{array}\right), \qquad
\p = \left(\begin{array}{ccc}  p_{\underline{2}1} & \sqrt{2} p_{\underline{2}2}  \\ \sqrt{2} p_{\underline{1}1} & -p_{\underline{2}1}  \end{array}\right),
\end{split}\end{equation}

Only in  this case $SP(4)/SU(2)$ is equivalent to  BD.I type symmetric space $SO(5)/(SO(3)\times SO(2))$ which is parametrized by a 3-component vector.

For $n=3$ each of the blocks $\q$ and $\p$ is parametrized by 6 matrix elements as follows:
\begin{equation}\label{eq:qp3}\begin{split}
\q = \left(\begin{array}{ccc} q_{1\underline{3}} & q_{1\underline{2}} & \sqrt{2} q_{1\underline{1}} \\ q_{2\underline{3}} & \sqrt{2} q_{2\underline{2}} & -q_{1\underline{2}} \\ \sqrt{2} q_{3\underline{3}} & -q_{2\underline{3}} & q_{1\underline{3}} \end{array}\right), \qquad \p = \left(\begin{array}{ccc} p_{\underline{3}1} & p_{\underline{3}2} & \sqrt{2}p_{\underline{3}3} \\ p_{\underline{2}1} & \sqrt{2}p_{\underline{2}2} & -p_{\underline{2}3} \\ \sqrt{2}p_{\underline{1}1} & -p_{\underline{1}2} & p_{\underline{1}3} \end{array}\right).
\end{split}\end{equation}

 Therefore $J$ and $ Q $ as $2n\times 2n$ dimensional matrices of the form:
\begin{equation}\label{eq:SpJQ}\begin{split}
Q(x,t) = \left(\begin{array}{cc} 0 &  \q \\  \p  &  0  \end{array}\right), \quad
 J = \frac{1}{2}\left(\begin{array}{ccc} \openone & 0 \\ 0 & -\openone   \end{array}\right).
\end{split}\end{equation}

The system of arising equations for the matrices $\p$ and $\q$ is formally the same as \eqref{eq:p-q_A3m}. The number of indepedent components of each matrix however is now $n(n+1)/2$, see for example \eqref{eq:qp2}, \eqref{eq:qp3}.
Noting that the Killing form is now given by \eqref{KillingC}, $\q=e^{-ix}\u,$ $ \p=e^{ix}\v, $ $\m=\u+i\u_x$ and $\n=\v-i\v_x,$ the Hamiltonians are
 \begin{equation} \label{C-H_1a}
\mathcal{H}_1=\frac{i}{2} \int \mathrm{tr}(\u_x \n) dx
 =-\frac{i}{2}  \int \mathrm{tr}(\v_x \m)dx;
\end{equation}
\begin{align} \label{C-H_2a}
\mathcal{H}_2 &=   -\frac{1}{2}\int \left( \mathrm{tr}( \u_{xx}\v ) + \mathrm{tr}( \m \v \u \v ) \right) dx \nonumber \\
&= -\frac{1}{2}\int \left( \mathrm{tr}( \u\v_{xx} ) +\mathrm{tr}( \n \u \v \u  ) \right) dx.
\end{align}
The first reduction with the Hermitian conjugation $\p=\pm \q^{\dagger},$ and the second nonlocal reduction reduction $\p(x,t)=\pm \q^{\dagger}(-x,-t)$ could be imposed on the equations. Further reductions are discussed in subsection \ref{c1r}.

\subsection{D.III symmetric space $SO^*(2n)/U(n)$}

The $so(2n)$ algebras are not isomorphic to any other simple Lie algebra only for $n\geq 4$. They may be understood as the set of all $2n\times 2n$ matrices $Y$, which satisfy
$$Y+ S_2Y^TS_2^{-1}=0,$$
where $S_2$ is the $2n\times 2n$ matrix. The explicit expressions for $S_2$ for even $n=2p$ and odd $n=2p+1$ are as follows:
\begin{equation}\label{eq:S2}\begin{aligned}
S_2^{(2p)} &= \sum_{k=1}^{2p}(-1)^{k+1}(E_{k\bar{k}} -E_{\bar{k},k}) = \left(\begin{array}{cc}
0 & s_2^{(2p)} \\ - s_2^{(2p)} & 0 \end{array}\right),  &\quad
s_2^{(2p)} &= \sum_{s=0}^{2p} (-1)^{s+1} E_{k,\underline{k}} ,\\
S_2^{(2p+1)} &= \sum_{k=1}^{2p+1}(-1)^{k+1}(E_{k\bar{k}} +E_{\bar{k},k}) = \left(\begin{array}{cc}
0 & s_2^{(2p+1)} \\  s_2^{(2p+1)} & 0 \end{array}\right), &\quad
s_2^{(2p+1)} &= \sum_{s=0}^{2p+1} (-1)^{s+1} E_{k,\underline{k}} ,
\end{aligned}\end{equation}
where $\bar{k}=2n+1-k $ and $\underline{k} = n+1 -k$.
Obviously $S_2^{-1}=S_2$ for all values of $n$, while $s_2^{-1}=(-1)^{n+1}s_2.$

The choice of Cartan involution in this case is related to $\vec{J}=\frac{1}{2} \sum_{s=1}^{n}{e_s}.$
The corresponding subsets $\Delta_0^+$ and $\Delta_1^+$ of the root system are
\begin{equation}\label{eq:D11}\begin{split}
 \Delta_0^+ = \{ e_k - e_j,\quad 1 \leq  k < j \leq n\}, \qquad \Delta_1^+ = \{ e_k + e_j, \quad 1 \leq k < j \leq n  \}.
\end{split}\end{equation}
Again, for $\alpha \in \Delta_0^+$ we have $(\alpha, \vec{J})=0$, while for $\beta \in \Delta_1^+$ we have $(\beta, \vec{J})=1$ and $a=1.$
Thus $\Delta_1^+$ contains $n(n-1)/2$ positive roots. The Cartan-Weyl generators can be represented in the form
\begin{equation}\label{eq:CW_D}\begin{aligned}
E_{e_k-e_j} &= E_{kj} - (-1)^{k+j} E_{\bar{j},\bar{k}}, &\quad E_{e_k+e_j} &= E_{k,\bar{j}} - (-1)^{k+j} E_{j,\bar{k}}, \\
H_{e_k} &= E_{k,k} - E_{\bar{k},\bar{k}}, & \quad E_{-\alpha} &= E_\alpha^T.
\end{aligned}\end{equation}
where $\bar{k} = 2n+1-k$. The Killing form is again $ \langle X,Y \rangle = \frac{1}{2}\text{tr}(XY). $ A generic potential is of the form:
\begin{equation} \label{expansion-gen'}
Q=\sum_{\alpha \in \Delta_1^+ } \left( \langle Q, E_{-\alpha}\rangle E_{\alpha}+\langle Q, E_{\alpha}\rangle E_{-\alpha} \right) = \left(\begin{array}{cc} 0 &  \q \\  \p  &  0  \end{array}\right),
\end{equation}
defining the local coordinates of the symmetric space has the above  block-off-diagonal form
 with $\p,\q$ being square $n\times n$ matrices satisfying (for both even and odd $n$)
 \begin{equation}\label{eq:so-pq}
\q+\s_{2}\q^T \s_{2}=0, \qquad \p+\s_{2}\p^T \s_{2}=0.\end{equation}
In other words, the components of the matrices $\p$ and $\q$ are not independent, they have $n(n-1)/2$ independent components, one for each root vector $\alpha \in \Delta_1^+$.

For $n=4$ for example, each of the blocks $\q$ and $\p$ is parametrized by 6 matrix elements  as follows
\begin{equation}\label{eq:U0}
\begin{split}
 q(x,t) = \left(\begin{array}{cccc}
q_{14} & q_{13} & q_{12} & 0 \\ q_{24} & q_{23} & 0 & q_{12} \\
q_{34} & 0 & q_{23} & -q_{13} \\ 0 & q_{34} & -q_{24} & q_{14} \\
\end{array}\right), \qquad p(x,t) &= \left(\begin{array}{cccc}
p_{14} & p_{24} & p_{34} & 0 \\ p_{13} & p_{23} & 0 & p_{34} \\
p_{12} & 0 & p_{23} & -p_{24} \\ 0 & p_{12} & -p_{13} & p_{14} \\
\end{array}\right).
\end{split}
\end{equation}
Moreover,
\begin{equation}\label{eq:D3J}\begin{split}
 J = \frac{1}{2}\left(\begin{array}{ccc} \openone & 0 \\ 0 & -\openone   \end{array}\right)
\end{split}\end{equation}
and the arising matrix equations also are in the form  \eqref{eq:p-q_A3m}, \eqref{eq:u-v_A3m} \eqref{eq:R1_A3}, \eqref{eq:R2_A3}. However, one has to keep in mind that $n\geq 4$ and $\p$ and $\q$ in this case have the structure as in (\ref{eq:U0}) which is totally different from the structures already mentioned before. Further reductions are discussed in subsection \ref{d3r}, see also \cite{NK-VG, VG-AS}.

\section{Reductions in a general form for the multicomponent FL equations }\label{s7}
In this Section we describe systematically the possible reductions of the multi-component FL equations derived previously. To this end we apply  Mikhailov's reduction group method \cite{Mi} to the multi-component FL Lax representations written in the following general form:
\begin{equation}\label{eq:gLM}\begin{split}
 L &\equiv i \frac{\partial }{ \partial x } + U(x,t,\lambda), \qquad  M\equiv i \frac{\partial }{ \partial t } + V(x,t,\lambda),
\end{split}\end{equation}
where $U(x,t,\lambda)$ and  $V(x,t,\lambda)$ are given in (\ref{eq:Lax}).

Mikhailov's  reduction group $G_R $ is a finite group which preserves the
Lax representation (\ref{eq:gLM}), i.e. it ensures that the reduction
constraints are automatically compatible with the evolution, defined by $[L,M]=0$. $G_R $ must
have two realizations: i) $G_R \subset {\rm Aut} \, \mathfrak{g} $ and ii) $G_R \subset {\rm Conf}\, \mathbb{C} $, i.e. as conformal mappings of the complex $\lambda $-plane. To each $g_k\in G_R $ we relate a reduction condition for the Lax pair as follows \cite{Mi}:
\begin{equation}\label{eq:2.1}
C_k(L(\Gamma _k(\lambda ))) = \eta _k L(\lambda ), \quad
C_k(M(\Gamma _k(\lambda ))) = \eta _k M(\lambda ),
\end{equation}
where $C_k\in \mbox{Aut}\; \mathfrak{g} $ and $\Gamma _k(\lambda )\in
\mbox{Conf\,} \mathbb{C} $ are the images of $g_k $ and $\eta _k =1 $ or $-1 $
depending on the choice of $C_k $. Since $G_R $ is a finite group then for
each $g_k $ there exist an integer $N_k $ such that $g_k^{N_k} =\openone
$. In all the cases below $ N_k=2 $ and the reduction group is isomorphic
to $\mathbb{Z}_2 $.

More specifically the automorphisms $C_k $, $k=1,\dots,4 $ listed above
lead to the following reductions for the matrix-valued functions  $U(x,t,\lambda)$ and  $V(x,t,\lambda)$ of the Lax representation:
\begin{eqnarray}\label{eq:U-V.a}
   \mbox{a)}& \qquad C_1(U^{\dagger}(\kappa _1(\lambda )))= U(\lambda ),
\qquad &C_1(V^{\dagger}(\kappa _1(\lambda )))= V(\lambda ), \\
\label{eq:U-V.b}
   \mbox{b)} & \qquad C_2(U^{T}(\kappa _2(\lambda )))= -U(\lambda ), \qquad
&C_2(V^{T}(\kappa _2(\lambda )))= -V(\lambda ), \\
\label{eq:U-V.c}
   \mbox{c)}& \qquad C_3(\bar{U}(\kappa _1(\lambda )))= -U(\lambda ), \qquad
&C_3(\bar{V}(\kappa _1(\lambda )))= -V(\lambda ), \\
\label{eq:U-V.d}
   \mbox{d)}& \qquad C_4(U(\kappa _2(\lambda )))= U(\lambda ), \qquad
&C_4(V(\kappa _2(\lambda )))= V(\lambda ),
\end{eqnarray}

Below we list only a few of the simplest Mikhailov type reductions. For more general reductions see \cite{VG-AS, NK-VG, VG-RI-AS, VG-tmf92}.

\subsection{Symmetric spaces of A.III  type $SU(n+m)/S(U(n)\times U(m))$.}

\begin{description}
  \item[Reduction a).] Let us define $C_1(U) = A_1 U A_1^{-1}$ and

  \begin{equation}\label{eq:A3.1}\begin{split}
    Q(x,t) = \left(\begin{array}{cc} 0 & \q \\ \p & 0   \end{array}\right),\qquad A_1 = \left(\begin{array}{cc} a_1 & 0 \\ 0 & b_1    \end{array}\right),
  \end{split}\end{equation}
  where $a_1$ and $b_1$ are $m\times m$ and $n\times n$ constant invertible matrices, $\q$ is $m\times n$ and $\p$ is $n\times m.$ Then the reduction conditions become
 \begin{equation}\label{eq:A32}\begin{split}
  A_1 J A_1^{-1} = J, \qquad  \kappa_1 A_1Q^\dagger A_1^{-1} = Q, \qquad \kappa_1^2 =1,
 \end{split}\end{equation}
  i.e.
\begin{equation}\label{eq:A33}\begin{split}
 \p = \kappa_1 b_1 \q^\dagger a_1^{-1}, \qquad  \q = \kappa_1 a_1 \p^\dagger b_1^{-1}, \qquad
 a_1 = a_1^\dagger , \qquad  b_1 = b_1^\dagger .
\end{split}\end{equation}
In particular, in the case of interest \eqref{eq:u-v1}  $m=1$ and we $a_1$ is just a constant scalar which could be scaled out to $a_1=1$ for the sake of simplicity. Then $ \p = \pm b_1 \q^\dagger  $ and therefore $ \v = \pm b_1 \u^\dagger  $ where $b_1$ is a constant Hermitian matrix. The reduced equation for the complex-valued vector $\u$ acquires the form
$$ i\u_{t}+ \u_{xx} - \u_{xt} \pm \big(2 (\u\cdot b_1 \u^\dagger) \u+(\u\cdot b_1 \u^\dagger)i \u_x +i(\u_x\cdot b_1 \u^\dagger) \u \big ) =0 .$$

The other two possibilities are not related to the vector-valued FL type equations, but rather to matrix-valued equations,  since these reductions require $n=m,$ in other words $\p$ and $\q$ are $n\times n$ matrix blocks:

 \item[Reduction b), $n=m$.] Let us define $C_2(U) = A_2 U A_2^{-1}$ where
  \begin{equation}\label{eq:A3.1'}\begin{split}
   A_2 = \left(\begin{array}{cc}  0 & a_2  \\ a_2^{-1} & 0    \end{array}\right),
  \end{split}\end{equation}
  and $a_2$ is $n\times n$ invertible matrix.
  Then the reduction conditions become
 \begin{equation}\label{eq:A34}\begin{split}
  A_2 J A_2^{-1} = -J, \qquad  \kappa_2 A_2 Q^T A_2^{-1} = -Q, \qquad \kappa_2^2 =1,
 \end{split}\end{equation}
  i.e.
\begin{equation}\label{eq:A35}\begin{split}
 \q = -\kappa_2 a_2 \q^T a_2, \qquad  \p = -\kappa_2 a_2^{-1} \p^T a_2^{-1}.
\end{split}\end{equation}
Note that $\p$ and $\q$ are not related between themselves.

 \item[Reduction c), $n=m$.] Let us define $C_3(U) = A_3 U A_3^{-1}$ where
  \begin{equation}\label{eq:A3.1''}\begin{split}
   A_3 = \left(\begin{array}{cc}  0 & a_3  \\ a_3^{-1} & 0    \end{array}\right),
  \end{split}\end{equation}
  where $a_3$ is $n\times n$ invertible matrix.
  Then the reduction conditions become
 \begin{equation}\label{eq:A36}\begin{split}
  A_3 J A_3^{-1} = -J, \qquad  \kappa_1 A_3\bar{Q} A_3^{-1} = -Q, \qquad \kappa_1^2 =1,
 \end{split}\end{equation}
  i.e.
\begin{equation}\label{eq:A37}\begin{split}
 \q = -\kappa_1 a_3 \bar{\p} a_3, \qquad  \p = -\kappa_1 a_3^{-1}\bar{ \q} a_3^{-1}, \qquad
 a_3 = \bar{a}_3.
\end{split}\end{equation}

\end{description}

\subsection{Symmetric spaces of BD.I  type $SO(2n+1)/(SO(2)\times SO(2n-1))$.}
We consider the simplest nontrivial realization of this symmetric space $SO(2n+1)/(SO(2)\times SO(2n-1))$. Then the potential $Q(x,t)$ and $J$ have the following $3\times 3$ block-structure:
\begin{equation}\label{eq:BD1}\begin{split}
 Q (x,t) = \left(\begin{array}{ccc} 0 & \q^T & 0 \\ \p & 0 & s_0 \q \\
 0 & \p^T s_0 &  0   \end{array}\right), \qquad J = \left(\begin{array}{ccc} 1 & 0 & 0 \\
 0 & 0 & 0 \\ 0 & 0 & -1  \end{array}\right),
\end{split}\end{equation} where $\p$ and $\q$ are $2n-1$-dimensional complex-valued vector-columns.

\begin{description}
\item[Reduction a).] Let   $C_1(U) = D_1 U D_1^{-1}$ where $ D_1\in SO(2n+1)$,
  \begin{equation}\label{eq:BD1'}\begin{split}
   D_1 = \left(\begin{array}{ccc} 1 & 0 & 0 \\ 0 & c_1 & 0 \\ 0 & 0 & 1 \end{array}\right),
  \end{split}\end{equation}
  where $c_1 \in SO(2n-1)$, i.e. $s_0 c_1 s_0 = c_1^{-1}$.
  Then the reduction conditions become
 \begin{equation}\label{eq:BD2}\begin{split}
  D_1 J D_1^{-1} = J, \qquad   \kappa_1 D_1 Q^\dagger D_1^{-1} = Q, \qquad \kappa_1^2 =1,
 \end{split}\end{equation}
  i.e.
\begin{equation}\label{eq:BD3}\begin{split}
 \p = \kappa_1 c_1 \bar{\q}, \qquad  \q = \kappa_1 s_0 c_1 s_0 \bar{\p}.
\end{split}\end{equation}

 \item[Reduction b).] Let us define $C_3(U) = D_3 U D_3^{-1}$ where $ D_3\in SO(2n+1)$,
  \begin{equation}\label{eq:BD3'}\begin{split}
   D_3 = \left(\begin{array}{ccc}  0 & 0 & 1 \\ 0 & a_3 & 0 \\ 1 & 0 & 0 \end{array}\right),
  \end{split}\end{equation}
  where $a_3$ is $n\times n$ invertible matrix.
  Then the reduction conditions become
 \begin{equation}\label{eq:BD4}\begin{split}
  D_3 J D_3^{-1} = -J, \qquad  \kappa_1 D_3\bar{ Q} D_3^{-1} = -Q, \qquad \kappa_1^2 =1,
 \end{split}\end{equation}
  i.e.
\begin{equation}\label{eq:BD5}\begin{split}
 \q = -\kappa_1 s_0a_3 \bar{\p} , \qquad  \p = -\kappa_1 a_3 s_0 \bar{\q} , \qquad
s_0 a_3^T s_0 =a_3^{-1}.
\end{split}\end{equation}

\end{description}

\subsection{Symmetric spaces of C.I  type $SP(2n)/SU(n)$}\label{c1r}

We recall that $Q(x,t) \in sp(n)$ and according to our definition of symplectic matrices we have, see \eqref{Sc}:
\begin{equation}\label{eq:C11'}\begin{split}
 Q(x,t) = \left(\begin{array}{cc} 0 & \q \\ \p & 0   \end{array}\right), \qquad
 S_1 Q + Q^T S_1 =0,  \qquad S_1 = \left(\begin{array}{cc} 0 & s_{12} \\ s_{21} & 0   \end{array}\right) ,\qquad s_{21}=-s_{12}^{-1}=(-1)^n s_{12}
\end{split}\end{equation}
where $\p$ and $\q$ are square $n\times n $ matrices and
$\q+\s_{12}\q^T \s_{12}^{-1}=0,$  $ \p+\s_{12}\p^T
\s_{12}^{-1}=0.$  The reductions below put additional conditions on $\q$ and $\p$.

\begin{description}
\item[Reduction a).] Let us define $C_1(U) = B_1 U B_1^{-1}$ where
  \begin{equation}\label{eq:C1.1}\begin{split}
   B_1 = \left(\begin{array}{cc} a_1 & 0 \\ 0 & b_1    \end{array}\right)
   \in SP(2n),
  \end{split}\end{equation}
  where $a_1$ and $b_1$ are $n\times n$ invertible matrices.
  Then the reduction conditions become
 \begin{equation}\label{eq:C12}\begin{split}
  B_1 J B_1^{-1} = J, \qquad  \kappa_1 B_1  Q^\dagger B_1^{-1} = Q, \qquad \kappa_1^2 =1,
 \end{split}\end{equation}
  i.e.
\begin{equation}\label{eq:C13}\begin{split}
 \p = \kappa_1 b_1 \q^\dagger a_1^{-1}, \qquad  \q = \kappa_1 a_1 \p^\dagger b_1^{-1}, \qquad
 a_1 = a_1^\dagger , \qquad  b_1 = b_1^\dagger .
\end{split}\end{equation}
We remind that the inner automorphisms of the algebra are always similarity  transformations by element of the corresponding group.
Note that the condition $B_1 \in SP(2n)$ means that the blocks $a_1$ and $b_1$ are related by $b_1= s_{12}(a_1^{-1})^T s_{12}^{-1}$.

 \item[Reduction b)] Let us define $C_3(U) = B_3 U B_3^{-1}$ where $C_3 \in SP(2n)$ and
  \begin{equation}\label{eq:A3.1A}\begin{split}
   B_3 = \left(\begin{array}{cc}  0 & a_3  \\ a_3^{-1} & 0    \end{array}\right)=B_3^{-1},
  \end{split}\end{equation}
  where $a_3$ is $n\times n$ invertible matrix.
  Then the reduction conditions become
 \begin{equation}\label{eq:C14}\begin{split}
  B_3 J B_3^{-1} = -J, \qquad   \kappa_1 B_3 \bar{Q} B_3^{-1} = -Q, \qquad \kappa_1^2 =1,
 \end{split}\end{equation}
  i.e.
\begin{equation}\label{eq:C15}\begin{split}
 \q = -\kappa_1 a_3 \bar{\p} a_3, \qquad  \p = -\kappa_1 a_3^{-1} \bar{\q} a_3^{-1}, \qquad
 a_3 = \bar{a}_3 .
\end{split}\end{equation}

\end{description}

\subsection{Symmetric spaces of  D.III type $SO^*(2n)/U(n)$} \label{d3r}

Here $Q(x,t) \in so(2n)$, $n\geq 4$ and according to our definition of orthogonal matrices we have:
\begin{equation}\label{eq:D31'}\begin{split}
 Q(x,t) = \left(\begin{array}{cc} 0 & \q \\ \p & 0   \end{array}\right), \qquad
 S_2 Q + Q^T S_2 =0.
\end{split}\end{equation}
where $\p$ and $\q$ are square $n\times n $ matrices and $S_2$ is defined in eq. (\ref{eq:S2}). The reductions below put additional conditions on $\q$ and $\p$, see also \cite{NK-VG, VG-AS}.
\begin{description}
\item[Reduction a).] Let us define $C_1(U) = D_4 U D_4^{-1}$ where $D_4 \in SO(2n)$ and
  \begin{equation}\label{eq:D1.1}\begin{split}
   D_4 = \left(\begin{array}{cc} a_4 & 0 \\ 0 & b_4    \end{array}\right),
  \end{split}\end{equation}
  In addition  $a_4$ and $b_4$ are $n\times n$ invertible matrices.
  Then the reduction conditions become
 \begin{equation}\label{eq:D12}\begin{split}
  D_4 J D_4^{-1} = J, \qquad  \kappa_1 D_4  Q^\dagger D_4^{-1} = Q, \qquad \kappa_1^2 =1,
 \end{split}\end{equation}
  i.e.
\begin{equation}\label{eq:D13}\begin{split}
 \p = \kappa_1 b_4 \q^\dagger a_4^{-1}, \qquad  \q = \kappa_1 a_4 \p^\dagger b_4^{-1}, \qquad a_4 = a_4^\dagger , \qquad  b_4 = b_4^\dagger .
\end{split}\end{equation}

\item[Reduction b).] Let us define $C_1(U) = D_5 U D_5^{-1}$ where $D_5 \in SO(2n)$ and
  \begin{equation}\label{eq:D1.2}\begin{split}
   D_5 = \left(\begin{array}{cc} 0 & a_5 \\ a_5^{-1} & 0    \end{array}\right),
  \end{split}\end{equation}
  In addition  $a_5$ is $n\times n$ invertible matrix.  Then the reduction conditions become
 \begin{equation}\label{eq:D12a}\begin{split}
  D_5 J D_5^{-1} = -J, \qquad  \kappa_1 D_5  \bar{Q} D_5^{-1} = -Q, \qquad \kappa_1^2 =1,
 \end{split}\end{equation}
  i.e.
\begin{equation}\label{eq:D13a}\begin{split}
 \p = -\kappa_1 a_5^{-1} \bar{\q} a_5^{-1}, \qquad  \q = -\kappa_1 a_5 \bar{\p} a_5, \qquad a_5 = \bar{a}_5 .
\end{split}\end{equation}

\end{description}

\section{Conclusions and discussion}\label{s8}
In conclusion, we obtained integrable systems of coupled equations associated to each irreducible Hermitian symmetric space. We have given illustrative examples with the ''classical'' series and their reductions. Further examples will be provided in future publications. The spectral theory, the Riemann-Hilbert approach for the underlying spectral problem, the study of the boundary value problems with the methods of the inverse scattering (see for example the review article \cite{P} and the references therein) are all challenging tasks, which remain to be addressed.
Another large area concerns the possible physical applications of the multi-component FL type equations. The physical models leading to such equations are not rare and here we provide only some indications in this direction.
The Schr\"odinger equation for Hamiltonians describing the dynamics of double-stranded DNA reduce to an equation that belongs to the following wider class of  equations for the 3-dimensional complex vector $\mathfrak{q}$ \cite{Qin}
\begin{equation}
i\mathfrak{q}_t+\mu_1\mathfrak{q}_{xx}+\mu_2 (\mathfrak{q}^\dagger \mathfrak{q})\mathfrak{q}+
i\varepsilon[\mu_3 \mathfrak{q}_{xxx}+\mu_4(\mathfrak{q}^\dagger \mathfrak{q})\mathfrak{q}_x +\mu_5(\mathfrak{q}^\dagger \mathfrak{q}_x)\mathfrak{q}+\mu_6(\mathfrak{q}^\dagger_x \mathfrak{q})\mathfrak{q}]=0
\end{equation}
where $\mu_k$ ($k=1,\ldots,6$) are constants related to the physical paramaters of the system (in general time-dependent, but here for simplicity we assume that they are constant) and $\varepsilon$ is a ''small-scale'' parameter; $(\mathfrak{q}^\dagger \mathfrak{q})=\sum_{k=1}^3 |\mathfrak{q}_k|^2$ is a scalar product. The variable change $$ \mathfrak{q}=\u-i\varepsilon \u_x$$ and the special choice of the parameters $ \mu_4=\mu_5 = \frac{1}{2}\mu_2 \mu_3,$ $ \mu_6=-\mu_2\mu_3 $ leads to the equation
\begin{equation} \label{example}
i\u_t+\mu_1\u_{xx}+ \varepsilon\mu_3 \u_{xt}+\frac{\mu_2}{2}\left[2 (\u^\dagger \u)\u - i\varepsilon \mu_3 \left((\u ^\dagger \u)\u _x +(\u^\dagger \u _x)\u \right)\right]=0
\end{equation} after neglecting all terms with $\varepsilon^2.$

On the other hand, rescaling of the variables in \eqref{eq:u-R1} with $n=3$ so that
$$ \partial_t \rightarrow \varepsilon^2\frac{\mu_3^2}{\mu_1}\partial_t, \quad \partial_x \rightarrow -\varepsilon \mu_3 \partial_x, \quad \u\rightarrow \sqrt{\frac{\mu_2}{2}} \u$$ leads to \eqref{example} as well. The proper discovery of the role of the FL equations in this area of course requires a lot of further efforts. We indicate also two other possible directions, and these are the settings where normally models of coupled multi-component NLS - type equations arise: in the studies of several layers of fluids, see for example \cite{AH} and the Spinor-Bose-Einstein condensates, \cite{GKV,Wad}.

\section*{Acknowledgements}
One of us (VSG) acknowledges support from the Bulgarian Science Foundation (grant NTS-Russia 02/101). The authors are thankful to two anonymous referees for their useful comments and suggestions.

\end{document}